%% file: Tech_Report.tex
\documentclass[a4paper, 11pt]{article}

\usepackage{graphicx}
\usepackage{color}
\usepackage{soul}
\usepackage{subfigure}
\usepackage{threeparttable}
\usepackage{caption}
\usepackage{epstopdf}
\usepackage{amsthm}
\usepackage{amssymb, amsmath, amsfonts,  bm}
\DeclareMathOperator*{\argmin}{\arg\!\min}

\usepackage[utf8]{inputenc} 
\usepackage{titlesec}
\titleformat{\paragraph}
{\normalfont\normalsize\bfseries}{\theparagraph}{1em}{}
\titlespacing*{\paragraph}
{0pt}{3.25ex plus 1ex minus .2ex}{1.5ex plus .2ex}
\usepackage[final]{epsfig}
\usepackage{float}
\usepackage{array}
\usepackage{amsmath,amssymb,latexsym}
\usepackage{multirow}

\usepackage[usenames, dvipsnames]{xcolor}
\usepackage{pgfplots, tikz}
\usepackage{tkz-berge}
\usetikzlibrary{plotmarks, patterns, arrows, automata, petri, topaths}
\usetikzlibrary{shapes.geometric}
\input{new_commands}

\usepackage{ifpdf}
\usepackage[english, onelanguage,ruled,vlined, longend, titlenotnumbered]{algorithm2e}
\usepackage{algorithmic}
\usepackage{url}
\usepackage[utf8]{inputenc}
\usepackage[T1]{fontenc}

\usepackage{siunitx}
\pgfplotsset{compat=1.12}
\usepackage[english, plain]{fancyref}
\newcommand*{\fancyrefsubseclabelprefix}{subsec}
\frefformat{plain}{\fancyrefsubseclabelprefix}{Subsection~#1}
\newcommand*{\fancyrefalglabelprefix}{alg}
\frefformat{plain}{\fancyrefalglabelprefix}{Algorithm~#1}
\frefformat{plain}{\fancyrefeqlabelprefix}{(#1)}
\usetikzlibrary{shapes.geometric}
\RequirePackage[l2tabu, orthodox]{nag}

\begin{document}

\title{\textbf{[Technical Report]}
\\ \emph{Multi-frequency calibration for DOA estimation with
distributed sensors}}

\bigskip

\author{Martin BROSSARD, Virginie OLLIER, Mohammed Nabil EL KORSO,
\\  R\'{e}my BOYER,
Pascal LARZABAL
\thanks{\scriptsize
M. Brossard is with Mines Paristech, France, V. Ollier is with
Safran Electronics and Defense, France,  M. N. El Korso is with Paris-Nanterre University, France,
R. Boyer is with University of Lille, France and P. Larzabal is with ENS Paris-Saclay, France. This work was supported by   ANR ASTRID project MARGARITA (ANR-17-ASTR-0015). Some of the contents have been partially published in \cite{BOEBL20}.}
}


\date{31 Jan 2020}

\maketitle



\begin{abstract}

In this work, we investigate direction finding in the presence of sensor gain uncertainties and directional perturbations for sensor array processing in a multi-frequency scenario. Specifically, we adopt a distributed optimization scheme in which coherence models are incorporated and local agents exchange information only between connected nodes in the network, i.e., without a fusion center. Numerical simulations highlight the advantages of the proposed parallel iterative technique in terms of statistical and computational efficiency.

\end{abstract}




\section{Introduction}
\label{Introduction}

Calibration and Direction-of-Arrival (DoA) estimation is a major issue in array processing \cite{stoica1990maximum,vorobyov2005maximum}. The latter has been studied in several applications, e.g., radar, sonar, satellite, wireless communication and radio interferometric systems \cite{van2013signal,godara1997application}, where we commonly use largely distributed sensors elements
aiming to achieve high resolution. In all these sensor network applications, calibration is required as some parameters are not exactly known due to imperfect instrumentation or propagation conditions \cite{ng1996sensor}.
    Let us note that calibration algorithms are distinguished by the presence \cite{ng1995active} or absence \cite{weiss1989array} of one or more cooperative sources, named calibrator sources. Indeed, prior source information can be available \cite{ng1996sensor} and consists mainly in the true/nominal directions and powers of calibrator sources (i.e., without any perturbation effects or antenna imperfections).
  Furthermore, most calibration algorithms are based on the least squares approach, with a sequential procedure updating each parameter alternatively \cite{van2013signal}. The least squares estimator is indeed equivalent to the Maximum Likelihood (ML) method under a (unrealistic) Gaussian noise model.

The aim of the proposed methodology here is to estimate successively the unknown sensor gains and phase errors, along with the calibrator and noise parameters, through minimization of a proper weighting cost function. In this work, uncertainties are estimated from the array covariance matrix, since dealing directly with time series data and operating on the signal domain quickly becomes computationally unfeasible for a large number of samples \cite{wijnholds2016blind}.
 The scenario under study is general but could be adapted to any practical application as in the radio astronomy context, where the number of parameters to estimate is tremendous and frequency bands are wide. 

In the multi-frequency scenario, a suboptimal way to perform calibration is to consider one wavalength bin at a time, with only one centralized processor, which has access to data in the whole available range of wavelengths.
In this work, we study an accelerated version based on the scalable form of the  Alternating Direction Method of Multipliers (ADMM) \cite{boyd2011distributed,kazemi2013clustered} with a specific network topology: there is no fusion center and agents exchange information only among themselves. The goal being to reduce the complexity in operation flow and signaling exchanging \cite{erseghe2012distributed,shi2014linear,mota2013d,ollierfast,yatawatta2016fine,yatawatta2012gpu}.
For estimation of the directional gains, the compressive sensing framework, especially the sparse representation method, is  well-adapted and has already been applied for source localization in fully and partially calibrated arrays \cite{malioutov2005sparse,steffens2014direction,haardt2014subspace,ollier2015article}.

The notation used through this paper is the following: $(.)^*$, $(.)^{T}$, $(.)^{H}$, $(.)^{\odot \alpha}$, $\Re(.)$ and $[.]_{n}$ denote, respectively, the complex conjugate, transpose, Hermitian operator, element-wise raising to $\alpha$, real part and the $n$-th element of a vector. The expectation operator is $\calE\{.\}$, $\otimes$, $\circ$ and $\odot$ denote, respectively,
 the Kronecker, the Khatri-Rao and the Hadamard product. The operator $\diag(.)$ converts a vector into a diagonal matrix, $\blkdiag(.)$ is the block-diagonal operator, whereas $\vectdiag(.)$ produces a vector from the main diagonal of a matrix and $\vect(.)$ stacks the columns of a matrix on top of one another. The operators $\left\|.\right\|_2$ and $\left\|.\right\|_{\sfF}$ refer to the $l_2$ and Frobenius norms, respectively. Finally, $\mathbf{I}_P$ is the $P \times P$ identity matrix and $ | \cdot|$ refers to the cardinality of a set.

\section{Model setup}
\label{setup}

Let us consider $Q$ emitting signal sources and $P$ sensor elements in the array. Each source direction $q \in \{1, \hdots,Q\}$ is defined by a 2-dimensional vector $\mathbf{d}_q = \left[d_q^{l},d_q^{m}\right]^{T}$, 
s.t., all nominal/true known directions, without any disturbances, are stacked in $\mathbf{D}^{\mathrm{K}} = \left[ \mathbf{d}_{1}^{\mathrm{K}} , \ldots , \mathbf{d}_{Q}^{\mathrm{K}} \right] \in \mathbb{R}^{2 \times Q}$. Propagation conditions induce wavelength dependent distortions, leading to apparent source directions $\mathbf{D}_{\lambda}=[\mathbf{d}_{1,\lambda}, \hdots,\mathbf{d}_{Q,\lambda}]$ different from the true ones.
Under the narrowband assumption, the array response matrix reads
$\mathbf{A}_{\mathbf{D}_{\lambda}} = \frac{1}{\sqrt{P}} \exp \left( -j \frac{2\pi}{\lambda} \boldsymbol{\Xi} \mathbf{D}_{\lambda} \right) $
in which $\boldsymbol{\Xi} =[\boldsymbol{\xi}_1,\hdots,\boldsymbol{\xi}_P]^T \in \mathbb{R}^{P \times 2}$ includes the known Cartesian coordinates describing each sensor location in the array, s.t., for $p \in \{1, \hdots,P\}$, $\boldsymbol{\xi}_p = [x_p,y_p]^T$. 
Therefore, the $P \times 1$ narrowband signals measured by all antennas is written as follows, for the \textit{n}-th time sample and wavelength $\lambda$,
\begin{equation}
\label{start}
\mathbf{x}_{\lambda}(n) = \mathbf{G}_{\lambda}\mathbf{A}_{\mathbf{D}_{\lambda}}\boldsymbol{\Gamma}_{\lambda}\mathbf{s}_{\lambda}(n) + \mathbf{n}_{\lambda}(n)
\end{equation}
where the undirectional antenna gains are collected in the complex diagonal matrix $\mathbf{G}_{\lambda}= \mathrm{diag}\{\mathbf{g}_{\lambda}\}\in \mathbb{C}^{P \times P}$ and the directional gain responses, assumed identical for all antennas, are modeled by the diagonal matrix $\boldsymbol{\Gamma}_{\lambda}\in \mathbb{C}^{Q \times Q}$. Finally, $\mathbf{s}_{\lambda}(n)  \sim \mathcal{CN}(\mathbf{0},\boldsymbol{\Sigma}_{\lambda})$ and $\mathbf{n}_{\lambda}(n)  \sim \mathcal{CN}(\mathbf{0},\boldsymbol{\Sigma}^{n}_{\lambda})$ are the i.i.d. calibrator source signal and additive Gaussian thermal noise vectors with their corresponding diagonal covariance matrices $\boldsymbol{\Sigma}_{\lambda} = \mathrm{diag}\{\boldsymbol{\sigma}_{\lambda}\} \in \mathbb{R}^{ Q \times Q}$ and $\boldsymbol{\Sigma}^n_{\lambda} = \mathrm{diag}\{\boldsymbol{\sigma}^n_{\lambda}\} \in \mathbb{R}^{P \times P}$, respectively.
From (\ref{start}), we deduce the following covariance matrix  \footnotemark
\begin{equation}
\label{model_cov}
\mathbf{R}_{\lambda}( \mathbf{p}_{\lambda})= \allowbreak \calE \left\lbrace \bfx_{\lambda} \bfx_{\lambda}^{\sfH} \right\rbrace = \mathbf{E}_{\mathbf{D}_{\lambda}} \mathbf{M}_{\lambda}\mathbf{E}^H_{\mathbf{D}_{\lambda}} + \boldsymbol{\Sigma}^{n}_{\lambda}
\end{equation}
where $
 \mathbf{E}_{\mathbf{D}_{\lambda}} = \mathbf{G}_{\lambda} \mathbf{A}_{\mathbf{D}_{\lambda}} \boldsymbol{\Sigma}^{1/2}_{\lambda}$
and $
 \mathbf{M}_{\lambda} = \boldsymbol{\Gamma}_{\lambda} \boldsymbol{\Gamma}^H_{\lambda} = \mathrm{diag}\{\mathbf{m}_{\lambda}\}.$
  \footnotetext{ As in \cite{martinjournalnew}, some commonly used assumptions are considered here to overcome scaling ambiguities, such as fixed phase for the first element and one reference source with fixed direction and directional gain/apparent power.
}
In this context, the calibration problem consists in estimating the parameter vector of interest
$ \mathbf{p} =  [ \mathbf{p}^T_{\lambda_1}, \hdots, \mathbf{p}^T_{\lambda_F} ]^T$
with $F$ the total number of available wavelengths  and $
 \mathbf{p}_{\lambda} = [ \mathbf{g}^T_{\lambda}, \mathbf{d}^T_{1,\lambda}, \hdots, \mathbf{d}^T_{Q,\lambda},\mathbf{m}^T_{\lambda}, \boldsymbol{\sigma}^{\mathrm{n}^T}_{\lambda}]^T$. To this end, we exploit sample covariance matrices $\hat{\mathbf{R}}_{\lambda}$, defined as
 $\hat{\mathbf{R}}_{\lambda} =  \frac{1}{N} \sum _{n=1}^N \mathbf{x}_{\lambda}(n) \mathbf{x}^H_{\lambda}(n)$
 for wavelength $\lambda$.

In estimation theory, the ML estimator is well-known for its statistical efficiency but not always easy to implement in practice. The Weighting Least Squares approach is an appropriate alternative  as it is asymptotically equivalent to the ML for a large number of samples $N$.
 Therefore, we wish to minimize the following local cost function, associated to wavelength $\lambda$
\begin{equation}
\label{cost_here}
\kappa_{\lambda}(\mathbf{p}_{\lambda}) = || \left ( \mathbf{R}_{\lambda} ( \mathbf{p}_{\lambda}) - \hat{\mathbf{R}}_{\lambda} \right) \odot \boldsymbol{\Omega}_{\lambda} ||^2 _{F}
\end{equation}
where $\boldsymbol{\Omega}_{\lambda} = (\boldsymbol{\sigma}_{\lambda}^{\rmn} \boldsymbol{\sigma}_{\lambda}^{\rmn^T})^{\odot - \frac{1}{2}}$. Most sources are assumed buried beneath the noise and antennas are identical in the array with negligible mutual coupling.
The aim of the designed calibration algorithm is to minimize the global cost function $\kappa(\bfp) = \sum_{\lambda \in \Lambda} \kappa_{\lambda}(\bfp_{\lambda})$ in a parallel and step-wise approach, with $\Lambda=\{\lambda_1,\hdots,\lambda_F\}$ the total set of available wavelengths. Usually, minimization is conducted w.r.t. one specific parameter while fixing the others in $ \mathbf{p}_{\lambda}$ \cite{martinjournalnew}.
Here, our approach is different: we propose an accelerated version where
estimation is performed directly w.r.t. the consensus (hidden) variables,  as described in Algorithm 1 and  detailed in the following.

\section{Description of the proposed estimator}

To achieve multi-frequency calibration in the sensor array, coherence is imposed along wavelength subbands for both directional and undirectional gains, by imposing available constraints or enforcing smooth variation.
The choice of the basis functions is motivated by the application under analysis and can be adapted accordingly.

\subsection{Coherence model for the undirectional antenna gains}
\label{cohe_undi}

 To impose coherence along subbands, we introduce a set of smooth wavelength dependent basis functions  and express the gains as linear combinations.
Let us define $\boalpha_{p}= [\alpha_{1,p}, \hdots, \alpha_{K_g,p} ]^T  \in \bbC^{K_g}$, the consensus vector for the $p$-th sensor  with unknown linear coefficients.
 Therefore, for  $p \in \{ 1, \hdots, P\}$ and $\lambda \in \Lambda$,
$ [\bfg_{\lambda}]_{p} = \sum_{k=1}^{K_g} b_{k, \lambda} \alpha_{k,p} = \bfb_{\lambda}^{\sfT} \boalpha_{p},$
in which $\bfb_{\lambda} = \left[b_{1, \lambda}, \ldots, b_{K_g, \lambda} \right]^{\sfT} \in \bbR^{K_g}$ stands for the polynomial terms, describing the variation of the undirectional gains w.r.t.~wavelength. For instance, we can consider the typical basis function $b_{k, \lambda} = \left(\frac{f-f_0}{f_0}\right)^{k-1}$ in which $f=c / \lambda$ is the studied frequency of interest with $c$ the speed of light and $f_0$ is the reference frequency \cite{martinjournalnew,yatawatta2015distributed}.
By stacking all vectors $\boalpha_{p}$, we obtain the global consensus vector
  $\boalpha = \left[\boalpha_{1}^{\sfT}, \ldots, \boalpha_{P}^{\sfT} \right]^{\sfT} \in \bbC^{P K_g},$
leading to
\begin{equation}
\label{express_gains_undi}
 \bfg_{\lambda} = \bfB_{\lambda} \boalpha,
\end{equation}
with $\bfB_{\lambda}= \left( \bfI_P \otimes \bfb_{\lambda}^{\sfT}   \right)$.

\subsection{Coherence model for the directional gains}

Similarly as for the undirectional gains, the coherence model is defined as follows: let us consider $\boalpha_{q} \in \bbR^{K_m}$, for $q \in \{ 1, \ldots, Q\}$, such that for $\lambda \in \Lambda$,
\begin{equation}
\label{express_gains_di_q}
 [\bfm_{\lambda}]_{q} =  \bfb_{\mathbf{m}_{\lambda}}^{\sfT} \boalpha_{\mathbf{m}_{q}},
\end{equation}
in which $\boalpha_{\mathbf{m}_{q}}$ is the vector of hidden variables for the $q$-th calibrator source, associated to directional gains $\bfm_{\lambda}$, while $\bfb_{\mathbf{m}_{\lambda}}$ is the corresponding basis vector. As in section \ref{cohe_undi}, all $\boalpha_{\mathbf{m}_{q}}$ are stacked in $
  \boalpha_{\mathbf{m}} = \left[\boalpha^T_{\mathbf{m}_{1}}, \ldots, \boalpha_{\mathbf{m}_{Q}}^{\sfT} \right]^{\sfT} \in \bbR^{Q K_m}$,
finally leading to
\begin{equation}
\label{express_gains_di}
 \bfm_{\lambda} = \bfB_{\mathbf{m}_{\lambda}} \boalpha_{\mathbf{m}}
\end{equation}
with $\bfB_{\mathbf{m}_{\lambda}}= \left( \bfI_Q \otimes \bfb_{\mathbf{m}_{\lambda}}^{\sfT}   \right)$. We assume identical behavior for all sources but the process can be straightforwardly adapted to different behavior. In \cite{martinjournalnew}, the directional gains in $\boGamma_{\lambda}$ were assumed inversely proportional to $\lambda$ but here the algorithm can be adjusted to any general existing models. 

\subsection{Distributed network with a fusion center}
\label{estim_alpha_z}

Dealing with large data volumes delivered by advanced sensor array systems requires computationally efficient calibration algorithms, with a huge number of unknowns to solve.
To improve both computational cost and estimation accuracy, distributed calibration has been proposed  by exploiting data parallelism across frequency. Contrary to a centralized hardware architecture which processes all frequency bands at a single location and is therefore computationally challenging, distributed optimization introduces more than one compute agents and analyzes the data simultaneously across smaller frequency intervals \cite{boyd2011distributed}. By distributing the total computations across the network, we gain a significant reduction in operational and energy cost and each agent receives information indirectly across the whole frequency range, thus improving the calibration accuracy.
To handle this, let us consider $Z$ computational agents disposed on a network. Each agent has access to some wavelengths $\lambda \in \Lambda_{z} =\{\lambda^{z}_{1}, \ldots, \lambda^{z}_{J_{z}}\} \subset \Lambda$. 
The corresponding unknown parameters in $\mathbf{p}$ are estimated locally and consensus is enforced among agents by imposing constraints in (\ref{express_gains_undi}) and (\ref{express_gains_di}).

To start with, let us focus on estimation of the undirectionnal sensor gains in section \ref{cohe_undi}.
We define $\boalpha^{z}$ as the local copy of the common optimization variable $\boalpha$ for the $z$-th agent and we note $\{\boalpha^{z}\}_{\calZ} = \{\boalpha^{1}, \ldots, \boalpha^{Z}\}$ the set of all $\boalpha^{z}$ in the network. Calibration is reformulated as the following constrained problem
\begin{equation}
 \begin{aligned}
  \hboalpha  = \argmin_{\boalpha, \{\boalpha^{z}\}_{\calZ}} \sum_{z=1}^{Z} \kappa^{z}\left(\boalpha^{z}\right) \ \ \text{subject to~} \boalpha^{z} = \boalpha \text{~for~} z \in \{1,\ldots,Z\}
 \end{aligned}
\end{equation}
where  $\kappa^{z}\left(\boalpha^{z}\right)$ is the cost function for the $z$-th agent, i.e., for $\lambda \in \Lambda_{z}$, which depends on the local variable $\boalpha^{z}$ and is associated to data  $\left\lbrace \hbfR_{\lambda}\right\rbrace_{\lambda \in \Lambda_{z}}$.
To solve this problem, we use the augmented Lagrangian, given by \cite{LiADMMcomplex}
$L \left( \{\boalpha^{z}\}_{\calZ}, \boalpha, \{\bfy^{z}\}_{\calZ} \right) = \sum_{z=1}^{Z}  \kappa^{z}\left(\boalpha^{z}\right)
  + \Re\left\{\bfy^{z \sfH} \left(\boalpha^{z} - \boalpha \right)\right\}  + \frac{\rho}{2} \left\| \boalpha^{z} - \boalpha \right\|_{2}^{2}  $
where $\{\bfy^{z}\}_{\calZ}$ are the $Z$ Lagrange multipliers and $\rho$ is the regularization term. We resort to the consensus ADMM in the scaled form by introducing the scaled dual variable $\bfu^{z}=\frac{1}{\rho}\bfy^{z}$ \cite{boyd2011distributed}. The three updates of the iterative algorithm are therefore given by
\begin{align}
 \boalpha^{z [t+1]} &= \argmin_{\boalpha^{z}} \kappa^{z}\left(\boalpha^{z}\right) + \frac{\rho}{2} \|\boalpha^{z}-\boalpha^{[t]}+\bfu^{z [t]}\|_{2}^{2} = \argmin_{\boalpha^{z}} \tL^{z}\left(\boalpha^{z},\boalpha^{[t]}, \bfu^{z [t]}\right)\label{eq:zetastep_first}  \\
 \boalpha^{[t+1]} &= \argmin_{\boalpha} \sum_{z=1}^{Z} \|\boalpha^{z [t+1]}-\boalpha+\bfu^{z [t]}\|_{2}^{2} \label{eq:zetastep} \\
 \bfu^{z [t+1]} &= \bfu^{z [t]} +  \left( \boalpha^{z [t+1]} - \boalpha^{[t+1]} \right)
\end{align}
where $t$ is the iteration counter.
Minimization (\ref{eq:zetastep}) leads to the following average, computed at the fusion center and sent to all agents in the network,
\begin{equation}
     \hboalpha = \frac{1}{Z} \sum_{z=1}^{Z} \left (\boalpha^{z} + \bfu^{z} \right ),
\end{equation}
from which the undirectional gains can be directly deduced with (\ref{express_gains_undi}).
The local minimization step in (\ref{eq:zetastep_first}) is the computationally most expensive one. To this end, we adopt an iterative approach and notice that the problem is separable w.r.t. each $\boalpha^{z}$, i.e., w.r.t. each agent. Let us assume $\boalpha^{z}$ and $(\boalpha^{z})^{\ast}$ as two independent variables \cite{salvini2014fast}. We then minimize $\tL^{z}\left(\boalpha^{z}, (\boalpha^{z})^{\ast},\boalpha, \bfu^{z}\right)$ w.r.t.  $\boalpha^{z}$, considering $(\boalpha^{z})^{\ast}$ as fixed and neglecting the diagonal elements in the cost function. In this case, the local cost function becomes separable w.r.t. the sub-vectors of $\boalpha^{z}$, i.e.,
 $\boalpha^{z} = \left[\boalpha_{1}^{z \sfT}, \ldots, \boalpha_{P}^{z \sfT} \right]^{\sfT},$
where $\boalpha_{p}^{z}$ is the local consensus vector for the $p$-th sensor at the $z$-th agent.
The following decompositions w.r.t. the sensor elements are also possible
\begin{align}
\label{decompose}
 \kappa^{z}(\boalpha^{z}) = \sum_{p=1}^{P} \kappa^{z}_{p}(\boalpha^{z}_{p})
\end{align}
and $ \tL^{z}\left(\boalpha^{z};\boalpha, \bfu^{z}\right) = \sum_{p=1}^{P} \tL^{z}_{p}\left(\boalpha^{z}_{p};\boalpha_p, \bfu^{z}_p\right)$ with
 $\tL^{z}_{p}\left(\boalpha^{z}_{p};\boalpha_p, \bfu^{z}_p\right) = \kappa^{z}_{p}(\boalpha^{z}_{p}) + \frac{\rho}{2} \|\boalpha^{z}_{p}-\boalpha_{p}+\bfu^{z}_{p}\|_{2}^{2}$
where $\kappa^{z}_{p}(\boalpha^{z}_{p})$ corresponds to the cost function for the $p$-th row of $\left\{\hat{\bfR}_{\lambda}\right\}_{\lambda \in \lambda_{z}}$, which only depends on $\boalpha^{z}_{p}$ since the remaining parameters are considered as fixed in this step. Let us define the operator $\calS_p(.)$, that converts to a vector the $p$-th row of a matrix and removes the $p$-th element of this selected vector. We also introduce the quantity $\bfRzero_{\lambda} = \bfA_{\bfD_{\lambda}} \boldsymbol{\Sigma}_{\lambda}\bfM_{\lambda}  \bfA^{\sfH}_{\bfD_{\lambda}}$ (reference source model) and the following vectors 
\begin{align}
 \hbfr^{\lambda}_{p} = \calS_p\left(\hbfR_{\lambda}\right) \odot \boomega^{\lambda}_{p},\ \ \ \  \ \
 \bfz^{\lambda}_{p} = \calS_p \Big(\bfRzero_{\lambda} \diag\left(\bfB_{\lambda} (\boalpha^{z})^{\ast}\right) \Big) \odot \boomega^{\lambda}_{p}
\end{align}
in which $\boomega^{\lambda}_{p} = \calS_p\left(\boOmega_{\lambda} \right) $.
In addition, let us consider the $J_{z} \times K_g$ matrix $
 \bfB^{z} = \left[\bfb_{\lambda^{z}_{1}}, \ldots, \bfb_{\lambda^{z}_{J_{z}}}\right]^{\sfT}$,
$\hbfr^{z}_{p} = \left[\hbfr^{\lambda^{z}_{1} \sfT}_{p}, \ldots, \hbfr^{\lambda^{z}_{J_{z}} \sfT}_{p}\right]^{\sfT} \in \mathbb{C}^{(P-1) J_z \times 1}$,
$\bfZ^{z}_{p} = \blkdiag\left(\bfz^{\lambda^{z}_{1}}_{p}, \ldots, \bfz^{\lambda^{z}_{J_{z}} }_{p}\right)  \in \mathbb{C}^{(P-1) J_z \times J_z}$ and
$\tbfZ^{z}_{p} = \bfZ^{z}_{p} \bfB^{z}$. 
We can thus write $\kappa^{z}_{p}(\boalpha^{z}_{p})$ in (\ref{decompose}) as
 $\kappa^{z}_{p}(\boalpha^{z}_{p}) = \left\| \hbfr^{z}_{p} - \tbfZ^{z}_{p} \boalpha^{z}_{p} \right\|_2^2$
and finally obtain the following estimate
\begin{equation}
 \hboalpha^{z}_{p} = \left( 2\tbfZ^{z \sfH}_{p} \tbfZ^{z}_{p} + \rho \bfI_{K_g} \right)^{-1} \left(2 \tbfZ^{z \sfH}_{p} \hbfr^{z}_{p} + \rho \left(\boalpha_{p} - \bfu^{z}_{p} \right)\right).
\end{equation}

\subsection{Distributed network with no fusion center}
\label{sec_no_fus}

We consider a specific formulation of the ADMM where every node in the network performs calibration locally and consensus is only reached with clearly identified neighbours without  fusion center \cite{erseghe2012distributed}. We note $\mathcal{N}_{z}$ the index set that corresponds to the neighbours of the $z$-th agent. The considered network architecture is exposed in Figure 1 where for example, $\mathcal{N}_{3} = \{2, 4\}$. We define the quantity $(\cdot)^{z,y}$ as the copy available at the $z$-th agent, transferred to the $y$-th agent.
In such context, the minimization problem becomes
\begin{equation}
\label{first_problem}
 \begin{aligned}
  \hboalpha  &= \argmin_{\{\boalpha^{z}, \bobeta^{z, y}, \forall y \in \calN_{z}  \}_{\calZ}} \sum_{z=1}^{Z} \kappa^{z}\left(\boalpha^{z}\right) \\
  &\text{subject to~} \boalpha^{z} = \bobeta^{z, y}, \ \  \bobeta^{y,z} = \bobeta^{z,y}, \forall y \in \calN_{z}, \text{~for~} z\in \{1,\ldots,Z\}
 \end{aligned}
\end{equation}
where the auxiliary variables $\bobeta^{z,y}$ impose consensus contraints on two neighboring agents and are meant to be local copies of $\boalpha$.
The decentralized strategy enables to  cooperatively minimize a sum of local objective functions, the final aim being to converge to a common  value, with fast convergence speed and good estimation performance \cite{erseghe2011fast}.
To obtain a more compact form of the problem in (\ref{first_problem}), we define
$\bobeta^{z} = \begin{bmatrix}
                               \{\bobeta^{z, y}\}_{y \in \calN_{z}}
               \end{bmatrix} \ \ \text{and} \ \
\bobeta = \begin{bmatrix}
                               \{\bobeta^{z}\}_{z\in\{1,\ldots, Z\}}
               \end{bmatrix} $,
leading to
\begin{equation}
 \begin{aligned}
  \hboalpha = \argmin_{\{\boalpha^{z}, \bobeta^{z}\}_{\calZ}} \sum_{z=1}^{Z} \kappa^{z}\left(\boalpha^{z}\right) \ \
  \text{subject to~} \bfH^{z} \boalpha^{z} = \bobeta^{z}, \text{~for~} z\in\{1,\ldots,Z\}, \ \
  \bobeta \in \calB
 \end{aligned}
\end{equation}
with 
 $\calB = \Big\{\bobeta |\bobeta^{z,y}=\bobeta^{y,z}, \forall y \in \calN_{z}, \text{~for~} z\in\{1,\ldots,Z\}\Big\}$
 and
  $\bfH^{z} = \bfun_{N_{z} \times 1} \otimes \bfI_{K_g P}$
where $N_{z}= |\calN_{z}|$.
As in section \ref{estim_alpha_z}, the scaled version of the ADMM leads to
\begin{align}
 \boalpha^{z [t+1]} &= \argmin_{\boalpha^{z}} \kappa^{z}\left(\boalpha^{z}\right) + \frac{\rho_{z}^{[t+1]}}{2} \|\bfH^{z} \boalpha^{z}-\bobeta^{z [t]}+\bfu^{z [t]}\|_{2}^{2} = \argmin_{\boalpha^{z}} \tL^{z}\left(\boalpha^{z}, \bobeta^{z [t]}, \bfu^{z [t]}\right)\label{eq:alpha_d_new}  \\
 \{\boldsymbol{\beta}^{z [t+1]}\}_{\calZ} &= \argmin_{\{\boldsymbol{\beta}^{z}\}_{\calZ} \in \calB} L \left( \{\boalpha^{z [t+1]}, \bobeta^{z}, \bfu^{z [t]}\}_{\calZ} \right)\label{eq:beta_d_new}  \\
 \bfu^{z [t+1]} &= \bfu^{z [t]} +  \left(\bfH^{z} \boalpha^{z [t+1]} - \bobeta^{z [t+1]} \right) \label{eq:u_d}
\end{align}
 and through decomposition of the problem in (\ref{eq:alpha_d_new}) w.r.t. sensor dependence,
 we obtain
\begin{equation}
\label{eq:alpha_z_estim}
 \hboalpha^{z}_{p} = \left( 2\tbfZ^{z \sfH}_{p} \tbfZ^{z}_{p} + \rho N_{z} \bfI_{K_g} \right)^{-1} \left(2 \tbfZ^{z \sfH}_{p} \hbfr^{z}_{p} + \rho \bfH_{p}^{z \sfH} \left(\bobeta^{z}_{p}-\bfu^{z}_{p} \right)\right)
\end{equation}
with
  $  \bfH^{z}_{p} = \bfun_{N_{z} \times 1} \otimes \bfI_{K_g}$. The selected variables $\bobeta^{z}_{p}$ and $\bfu^{z}_{p}$ are obtained from $\bobeta^{z}$ and $\bfu^{z}$ via an appropriate selection matrix.
After considering the projection onto $\calB$ and denoting the messages passed between the agent as
\begin{equation}
\label{eq:gamma}
 \bogamma^{z [t+1]} = \begin{bmatrix} \{\bogamma^{z,y [t+1]}\}_{y \in \calN_{z}}
             \end{bmatrix} = \bfH^{z} \boalpha^{z [t+1]} + \bfu^{z [t]},
\end{equation}
we solve (\ref{eq:beta_d_new}) thanks to
\begin{align}
\label{eq:beta}
 \bobeta^{z,y [t+1]} = \frac{1}{2} \left(\bogamma^{y,z [t+1]} + \bogamma^{z,y [t+1]} \right).
\end{align}
The steps of the proposed distributed method for calibration of sensor gains are exposed in Algorithm 1.2.

\subsection{Estimation of directional gains}

In this section, we describe the part of the algorithm dedicated to the estimation of DoA $\mathbf{D}_{\lambda}$ and directional gains $ \bfm_{\lambda} $, for fixed sensor gains, with a sparse and distributed implementation. Assuming a sparse observed scene, we define dictionaries of steering matrices for $q \in \{1,\ldots Q\}$ and $\lambda \in \Lambda$, as
$ \tbfA_{\lambda} = \left[ \tbfA_{1, \lambda}, \ldots ,\tbfA_{Q, \lambda} \right] \in \bbC^{P \times N_g},$
where $N_g = \sum_{q=1}^{Q} N_q$ denotes the total number of directions on the grid. The sparse vectors in
 $\tbfm_{\lambda} = \left[ \tbfm_{1, \lambda}^{\sfT}, \ldots ,\tbfm_{Q, \lambda}^{\sfT} \right]^{\sfT} \in \bbR^{N_g},$
contain the corresponding squared direction dependent gains. 
The covariance model is rewritten as
 $ \bfR_{\lambda} = \tbfE_{\lambda}  \tbfM_{\lambda} \tbfE_{\lambda}^{\sfH}  + \boSigman_{\lambda} \text{,}$
in which $\tbfM_{\lambda} = \diag(\tbfm_{\lambda})=\left(\bfI_{N_{g} } \otimes \bfb^T_{\lambda}\right)\blkdiag \left(\boldsymbol{\alpha}_1,\ldots, \boldsymbol{\alpha}_{N_g}\right)$,  $\tbfE_{\lambda} = \bfG_{\lambda} \tbfA_{\lambda} \tboSigma^{\frac{1}{2}}_{\lambda}$ and $\tboSigma_{\lambda} = \blkdiag \left(\bfI_{N_1 } \left[\bosigma_{\lambda}\right]_1,\ldots, \bfI_{N_Q }\left[\bosigma_{\lambda}\right]_Q\right)$.
To handle the DoA estimation and satisfy both sparsity and positivity requirements, we use the Distributed Iterative Hard Thresholding (IHT) \cite{blumensath2010normalized,patterson2014distributed}.
But contrary to \cite{martinjournalnew},  the following hard-thresholding operator
$\calH_1 \left(  \sum_{\lambda \in \Lambda} \left(\cbfV_{\lambda}^{q \sfT} \chbfr_{\lambda}^{q}\right)^{\odot 2} \right)$
is considered to provide access to the DoA of the $q$-th source, and a first estimate of the directional gain $\check{\mathbf{m}}^z_{q}$. The quantity $(\cdot)^q$ refers to the $q$-th column of a matrix, the expression $\check{(\cdot)}$ discards the elements corresponding to the diagonal of $\hat{\mathbf{R}}_{\lambda}$ and the hard thresholding operator $\calH_s(.)$ keeps the $s$-largest components and sets the remaining entries equal to zero. %
Finally, thanks to (\ref{express_gains_di_q})
and dealing  with the consensus variables as in section \ref{sec_no_fus}, the minimization problem becomes
\begin{equation}
 \begin{aligned}
  \hboalpha_{\mathbf{m}_{q}}  = \argmin_{\{\boalpha^{z,z}_{\mathbf{m}_{q}}, \{\boalpha^{z,y}_{\mathbf{m}_{q}}\}_{y \in \calN_{z}} \}_{\calZ}} \sum_{z=1}^{Z} \eta^{z}_{q}\left(\boalpha^{z,z}_{\mathbf{m}_{q}}\right) \ \
  \text{subject to~} \boalpha^{z,z}_{\mathbf{m}_{q}}= \boalpha^{y,z}_{\mathbf{m}_{q}}, \forall y \in \calN_{z}, \text{~for~} z \in \{1,\ldots,Z\}
 \end{aligned}
\end{equation}
where we benefit from the previous hard-thresholding estimate to define
 $\eta^{z}_{q}\left(\boalpha_{\mathbf{m}_{q}} \right) = \sum_{\lambda \in \Lambda_{z}} \left\| \cm_{q,\lambda} - \bfb^T_{\mathbf{m}_{\lambda}}  \boalpha_{\mathbf{m}_{q}} \right\|_{2}^{2}
 = \left\| \cbfm^{z}_{q} - \bfB^{z}_{\mathbf{m}} \boalpha_{\mathbf{m}_{q}} \right\|_{2}^{2}$
with $\cbfm^{z}_{q} = [\cm_{q,\lambda_{1}^{z}}, \ldots, \cm_{q,\lambda_{J_{z}}^{z}}]^T$ and $\bfB_{\mathbf{m}}^{z} = \left[\bfb_{\mathbf{m}_{\lambda_{1}^{z}}}, \ldots, \bfb_{\mathbf{m}_{\lambda_{J_{z}}^{z}}}
\right]^{\sfT}$.
As previously, we impose consensus between neighbours thanks to some auxiliary variables
but due to lack of space, we only present here the resulting local update for $\boalpha^{z}_{\mathbf{m}_{q}}$,
\begin{align}
\label{eq:estim_alpha_m}
 \hboalpha^{z}_{\mathbf{m}_{q}} &= \left( 2 \bfB_{\mathbf{m}}^{z \sfT} \bfB_{\mathbf{m}}^{z} + \rho^{z} \bfH_{\mathbf{m}}^{z^{\sfT}} \bfH_{\mathbf{m}}^{z} \right)^{-1} \left( 2 \bfB_{\mathbf{m}}^{z \sfT} \cbfm^{z}_{q} + \rho^{z} \bfH_{\mathbf{m}}^{z^{\sfT}} \left( \bobeta_{\mathbf{m}}^{z} - \bfu_{\mathbf{m}}^{z} \right) \right)
\end{align}
where $\bfH^{z}_{\mathbf{m}} = \bfun_{N_{z} \times 1} \otimes \bfI_{K_m \times K_m}.$
From $\hboalpha^{z}_{\mathbf{m}_{q}}$, we obtain an estimate of $[\bfm_{\lambda}]_q$ and process the next source, as shown in Algorithm 1.3. 

%
%
%

\section{Numerical simulations}

In order to evaluate the method, we consider realistic simulations for the radio astronomy context where the new generation of phased array systems such as the Low Frequency Array (LOFAR) and the Square Kilometre Array (SKA) requires the development of new advanced signal processing techniques for calibration purpose \cite{van2013signal,ollier2016relaxed}.  Indeed, lack of calibration leads to dramatic effects and distortions in the reconstructed images.
We consider $P=60$ antennas spread over a five-armed spiral \cite{wijnholds2004sky,ollierjournal}, which corresponds to the LOFAR's Initial Test Station. Let us assume a sky model with $Q=3$ strong calibrator sources and $Q^{\OUT}=8$ weak unknown sources in the background. 
The reference frequency $f_0 $ is set to $ \SI{30}{MHz}$ and we consider frequencies ranging from $\SI{29.6}{MHz}$ to $\SI{30.4}{MHz}$, with $Z=3$ agents in the network and $N_{z}=2$. The polynomial orders are chosen as $K_g=K_m=3$.
 The consensus variables $\boldsymbol{\alpha}$ and $\boldsymbol{\alpha}_{\mathbf{m}}$ are initialized as zeros and the squared directional gains are generated thanks to power law functions $(\lambda / \lambda_0)^{k-1}$ for $k \in \{1,\hdots,K_m\}$.

\subsection{Influence of the number of frequency channels}

First of all, we investigate the statistical performance of the proposed distributed algorithm as a function of the number of samples $N$ or the Signal-to-Noise Ratio (SNR). The SNR is defined as the ratio between the sum of apparent powers for all $Q$ sources  and the noise power. Results are averaged for $100$ Monte-Carlo runs. In Figure 2, we plot the three following cases: $F=3$ and each agent handles one frequency, i.e., $J_z=1$ (green curve), $F=9$ with $J_z=3$ (blue curve) and $F=27$ with $J_z=9$ (red curve). In Figure 2 (a), we plot the Root Mean Square Error (RMSE) as a function of $N$ for the undirectional gains $\mathbf{g}_{\lambda}$, defined as 
  $  \epsilon^{\bfg_{\lambda}}_{\RMSE} = \frac{1}{\sqrt{PF}} \sum_{\lambda \in \Lambda} \left\| \bfB_{\lambda}\hboalpha - \bfB_{\lambda}\boalpha \right\|_{2},$
for fixed SNR = $- 36$ dB.
A similar figure is presented in Figure 2 (b), for the source directions $\mathbf{D}_{\lambda}$, as a function of the SNR and fixed $N=2^{8}$.
We illustrate the performance by comparing with the mono-calibration scenario where each agent handles one single frequency, independently.  We notice that mono-calibration is clearly improved, by using a distributed procedure where the whole information is flowing through the entire network.

\subsection{Influence of the network architecture}

We aim to show the advantages of the proposed distributed network with no fusion center and only exchange of local information between neighboring agents, in terms of complexity.
With similar number of iterations in all loops of the algorithm, different estimation performance are attained in Figure 2 (a) while similar RMSE is reachable in Figure 2 (b) but with an additional computational cost if there is a fusion center (an increase of at least a factor 5 in computing time).

\subsection{Convergence analysis}

We illustrate the convergence behavior of the proposed algorithm by analyzing the following residuals as function of the iteration number.
Depending on the iteration in Algorithm 1, we plot the  primal residual as a function of the iteration number of Algorithm 1.2, defined as 
$\epsilon_p^{[t]} = \frac{1}{\sqrt{PK_gZN_z}} \sum_{z=1}^{Z} \left\| \bfH^{z}\boalpha^{z[t]} - \bobeta^{z[t]} \right\|_{2} .$
Likewise, we also study the different estimates between agents through
 $   \epsilon^{[t]}_{\mathrm{DIFF}} = \frac{1}{\sqrt{PK_gZ(Z-1)}} \sum_{z,z'=1}^{Z} \left\|  \boalpha^{z [t]} - \boalpha^{z' [t]}   \right\|_{2} .$
Similar statistical behavior can be obtained for corresponding residuals in Algorithm 1.3

\section{Conclusion}

In this work, we proposed an iterative algorithm for parallel calibration, applied in a general context of sensor array processing: complex electronic gains are imprecisely known and propagation disturbances lead to deviations in the source locations.  In order to reduce the communication overhead, the specific variation of parameters across wavelength is exploited in a distributed network with no fusion center and local exchange of information between adjacent connected nodes.
The two main steps of the algorithm are based on the scalable form of the ADMM and distributed IHT procedures.
 We highlighted the effectiveness and time efficiency of the proposed method using simulated data, even in the presence of non-calibrator sources at unknown directions.




\begin{algorithm}[h]
 \KwIn{$\big\lbrace \hbfR_{\lambda}\big\rbrace_{\lambda \in \Lambda}, \bfD^{\TRUE}, \eta_{\bfp}$\;}
 \textbf{Initialize:} set $i = 0, \big\lbrace \bfg_{\lambda} = \bfg_{\lambda}^{[0]},\bfD_{\lambda}=\bfD^{\TRUE}, \bfm_{\lambda} = \bfm_{\lambda}^{[0]}, \boOmega_{\lambda} = \bfun_{P \times P} \big\rbrace_{\lambda \in \Lambda}$\;
 \Repeat{$\left\| \bfp^{[i-1]}-\bfp^{[i]} \right\|_2 \le \left\| \bfp^{[i]} \right\|_2 \eta_{\bfp}$}{
 \nl $i = i+1$\;
 \nl Estimate in parallel $\big\lbrace \bfg_{\lambda}^{[i]} \big\rbrace_{\lambda \in \Lambda}$ \label{lst:line:g} with \textbf{Algorithm 1.2}\;
 \nl Estimate in parallel $\big\lbrace \bfD^{[i]}_{\lambda}, \bfm_{\lambda}^{[i]}, \bosigma^{\rmn[i]}_{\lambda} \big\rbrace_{\lambda \in \Lambda}$ \label{lst:line:doa} with \textbf{Algorithm 1.3}\;
 \nl Update locally $\big\lbrace \boOmega_{\lambda}^{[i]}  \big\rbrace_{\lambda \in \Lambda}$\;}
 \KwOut{$\hat{\bfp} = \big[\bfp_{\lambda_1}^{[i]\sfT}, \ldots, \bfp_{\lambda_F}^{[i]\sfT} \big]^{\sfT}$\;}
 \caption[]{Proposed calibration algorithm}
\end{algorithm}

\begin{algorithm}[ht]
\SetAlgoRefName{1.2}
 \KwIn{$\big\lbrace \hbfR_{\lambda} \big\rbrace_{\lambda \in \Lambda}, \bfp^{[i-1]}, \eta_{\boalpha}$\;}
 \textbf{Initialize:} set $t = 0, \boalpha^{z} = \boalpha^{[i-1]}$, $\bfRzero_{\lambda} = \bfA_{\bfD^{[i-1]}_{\lambda}}  \boldsymbol{\Sigma}_{\lambda} \bfM^{[i-1]}_{\lambda}   \bfA^{\sfH}_{\bfD^{[i-1]}_{\lambda}}$\;
  \While{stop criterion unreached}{
   \nl $t = t+1$ \;
   \nl Estimate locally $\boalpha^{z [t]}$ with \textbf{Algorithm 1.2.2}\;
   \nl Calculate locally $\bogamma^{z [t]}$ with (\ref{eq:gamma})\;
   \nl $\Rightarrow$ Broadcast values $\bogamma^{z,y [t]}$ to region $y \in \calN_{z}$\;
   \nl $\Leftarrow$ Receive values $\bogamma^{y,z [t]}$ from region $y \in \calN_{z}$\;
   \nl Estimate locally $\bobeta^{z [t]}$ with (\ref{eq:beta}) \;
   \nl Update locally $\bfu^{z [t]}$ with (\ref{eq:u_d})\;
   \nl Update possibly $\rho^{[t]}_{z}$ with \cite{boyd2011distributed,ghadimi2015optimal}\;
  }
 \caption{Distributed estimation of consensus variables for undirectional gains}
\end{algorithm}

\begin{algorithm}[ht]
\SetAlgoRefName{1.2.2}
 \KwIn{$\big\lbrace \hbfR_{\lambda}, \bfRzero_{\lambda},\big\rbrace_{\lambda \in \Lambda_{z}}, \boalpha^{z [t-1]}, \bobeta^{z [t-1]}, \bfu^{z [t-1]}, \eta_{\boalpha^{z}}$;}
 \textbf{Initialize:} set $t^{z} = 0, \boalpha^{z} = \boalpha^{z [t-1]}$\;
   \While{$\left\| \boalpha^{z [t^{z}-1]}-\boalpha^{z [t^{z}]}\right\|_2 \ge \left\| \boalpha^{z [t^{z}]} \right\|_2 \eta_{\boalpha^{z}}$}{
    \nl $t^{z} = t^{z}+1$ \;
    \For{$p \in \{ 1, \ldots, P\}$}{
        \nl Update $\tbfZ^{z}_{p}$\;
        \nl Estimate $\hboalpha^{z [t^{z}]}_{p}$ with (\ref{eq:alpha_z_estim}) \;
       \nl Update $(\hboalpha^{z}_{p})^{\ast}$ \;
     }
  }
 \KwOut{$\hboalpha^{z} = \boalpha^{z [t^{z}]}$\;}
 \caption{local estimation of $\boalpha^{z}$}
\end{algorithm}

\begin{algorithm}[ht]
\SetAlgoRefName{1.3}
 \KwIn{$\big\lbrace\hbfR_{\lambda}\big\rbrace_{\lambda \in \Lambda}, \bfp^{[i-1]}, \eta_{\boalpha_{\mathbf{m}}},\eta_{\bfD}, \hbfr_{\lambda} = \vect\left(\hbfR_{\lambda} \odot \boOmega_{\lambda}\right)\text{,}$ $ \bfV_{\lambda} = \left ( \boSigma^{\rmn}_{\lambda} \right )^{-\frac{1}{2}} \tbfE^{*}_{\lambda} \otimes \left ( \boSigma^{\rmn}_{\lambda} \right )^{-\frac{1}{2}} \tbfE_{\lambda}$\;}
 \textbf{Init:} set $k = 0$, $\bfg_{\lambda}^{[k]} = \bfg_{\lambda}^{[i]}$, $\bfM_{\lambda}^{[k]} = \bfM^{[i-1]}_{\lambda}$, $\bfD_{\lambda}^{[k]} = \bfD^{[i-1]}_{\lambda}, \bosigma_{\lambda}^{\rmn [k]} = \bosigma_{\lambda}^{\rmn[i-1]}$\;
 \While{$\left\| \boalpha_{\mathbf{m}}^{[k-1]}- \boalpha_{\mathbf{m}}^{[k]} \right\|_2 \ge \left\| \boalpha_{\mathbf{m}}^{[k]} \right\|_2 \eta_{\boalpha_{\mathbf{m}}}$ and  $\sum_{\lambda \in \Lambda} \left\| \bfD_{\lambda}^{[k-1]}- \bfD_{\lambda}^{[k]} \right\|_{\sfF} \ge  \eta_{\bfD}$}{
 \nl $k = k+1$\;
 \For{$q\in \{1,\ldots, Q\}$}{
  \ForEach{$\sfA_z, z \in \{ 1, \ldots, Z\}$}{
   \ForEach{$\lambda \in \Lambda_{z}$}{
    \nl Calculate locally the residual $\chbfr_{\lambda}^{q}$ as indicated in \cite{martinjournalnew}\;
    }
   \nl Compute $\calH_1 \left(  \sum_{\lambda \in \Lambda} \left(\cbfV_{\lambda}^{q \sfT} \chbfr_{\lambda}^{q}\right)^{\odot 2} \right)$ as indicated in \cite{martinjournalnew}\;
   }
  \nl Deduce $\hbfd_{q,\lambda}$ and $\check{\mathbf{m}}^z_{q}$ for each wavelength and agent\;
  \nl Estimate $[\mathbf{m}_{\lambda}]_q$ with similar procedure than \textbf{Algorithm 1.2} and (\ref{eq:estim_alpha_m})\;
  }
 }
 \nl Estimate locally $\bosigma_{\lambda}^{\rmn}$ as indicated in \cite{martinjournalnew} \;
 \caption{Distributed estimation of $\left\lbrace \bfm_{\lambda}, \bfD_{\lambda}, \bosigma_{\lambda}^{\rmn}\right\rbrace_{\lambda \in \Lambda}$ \label{alg:iht}}
 \KwOut{$\big\lbrace\hbfm_{\lambda}, \hbfD_{\lambda}, \hbosigma_{\lambda}^{\rmn}\big\rbrace_{\lambda \in \Lambda}$ \;}
\end{algorithm}

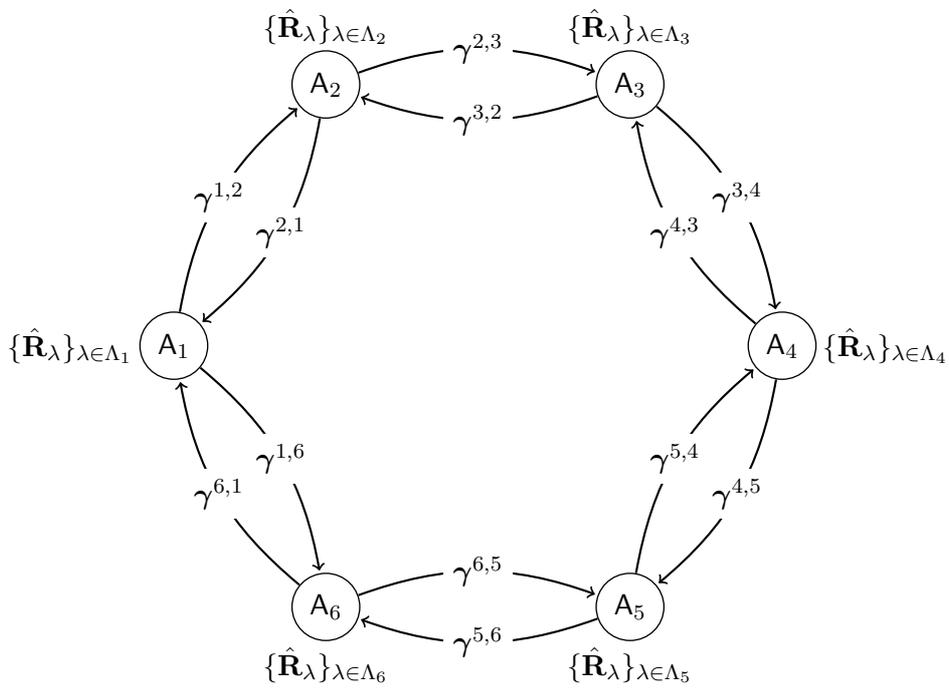
\begin{figure}
    \centering
    \input{tikz_dessin.tex}
    \caption*{Figure 1: Example of distributed network with no fusion center.}
    \label{fig:dessin}
\end{figure}

 \begin{figure}[t!]
 \centering
\subfigure[]{\label{fig:res2-a}\includegraphics[width=8cm, height=7cm]{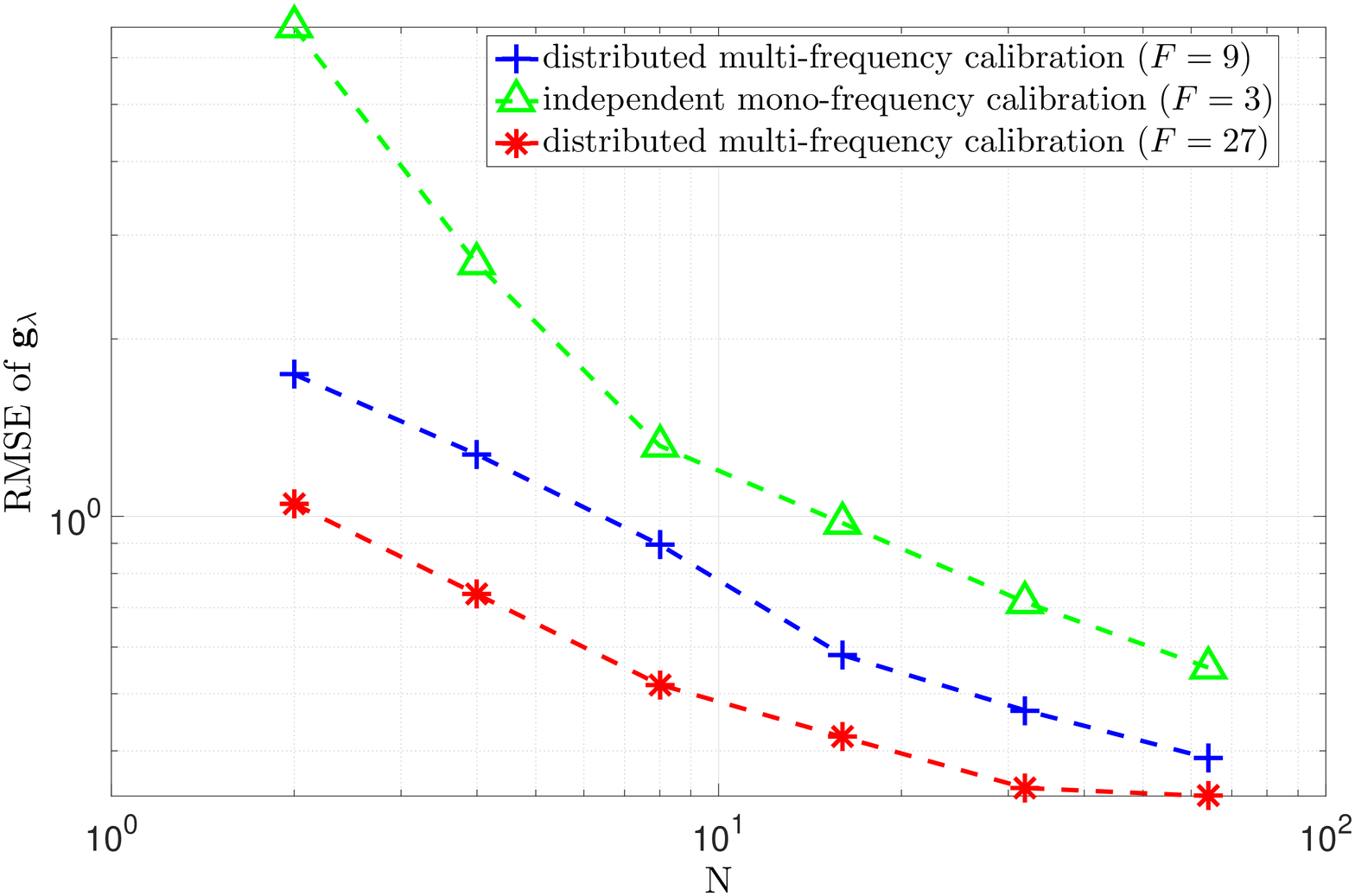}} \hspace{0.5pt} %
\subfigure[]{\label{fig:res2-b}\includegraphics[width=8cm, height=7cm]{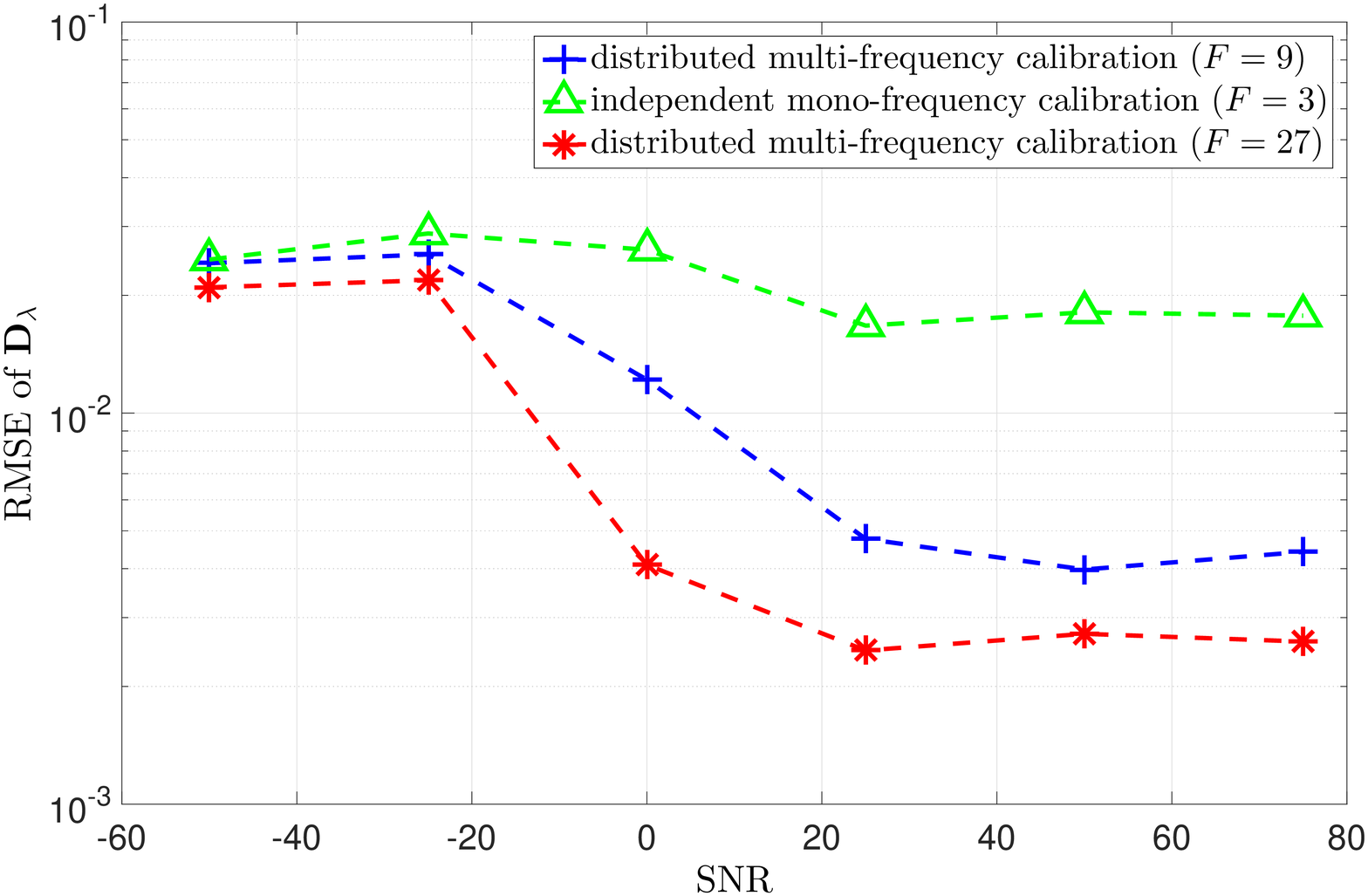}}
\caption*{Figure 2: (a) RMSE on the undirectional gains as function of the number of samples $N$, (b) RMSE on the apparent source directions as function of the SNR. }
\end{figure}

\begin{figure}[t!]
 \centering
\subfigure[]{\label{fig:res2-a}\includegraphics[width=8cm, height=7cm]{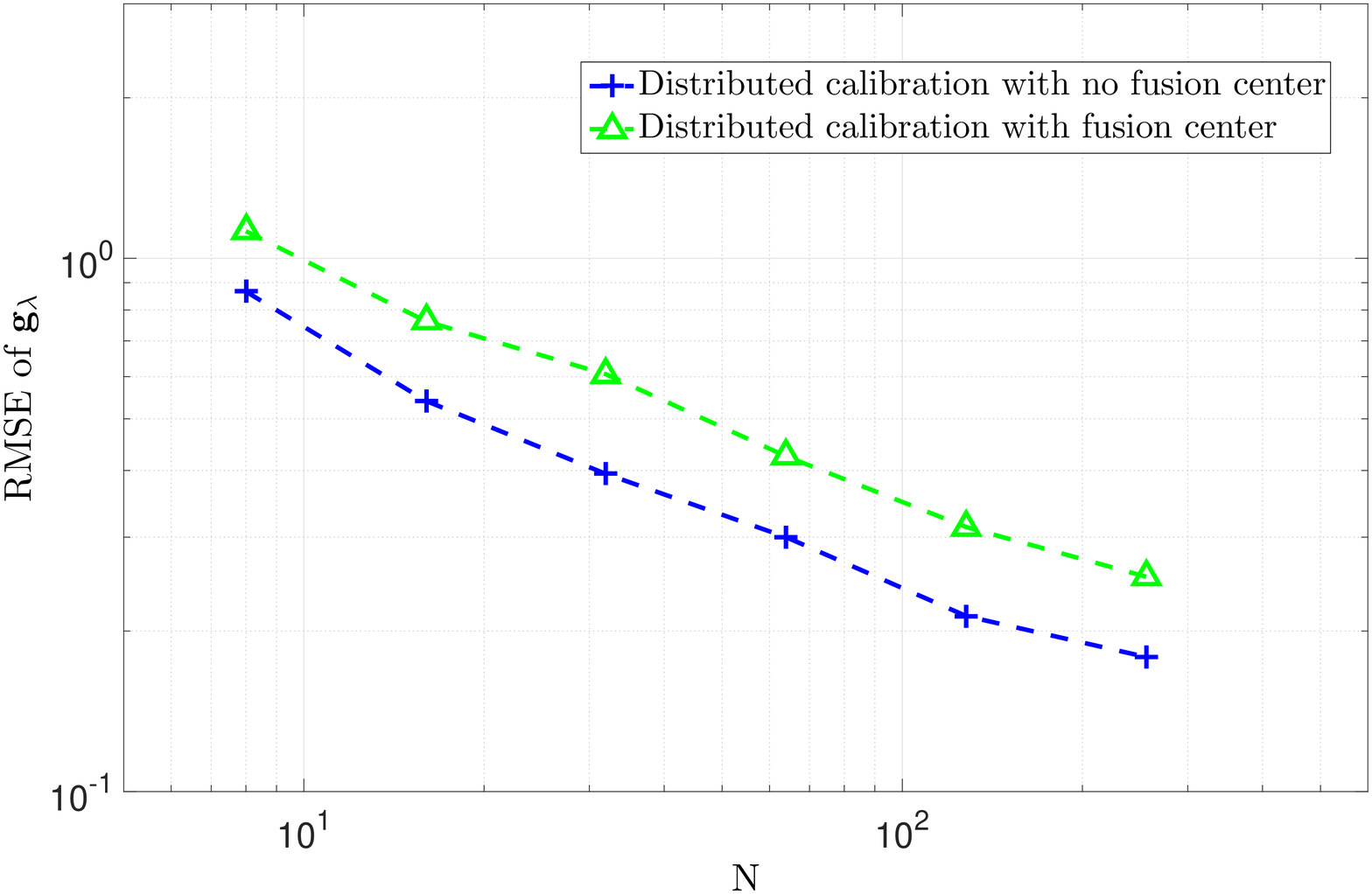}} \hspace{0.5pt} %
\subfigure[]{\label{fig:res2-b}\includegraphics[width=8cm, height=7cm]{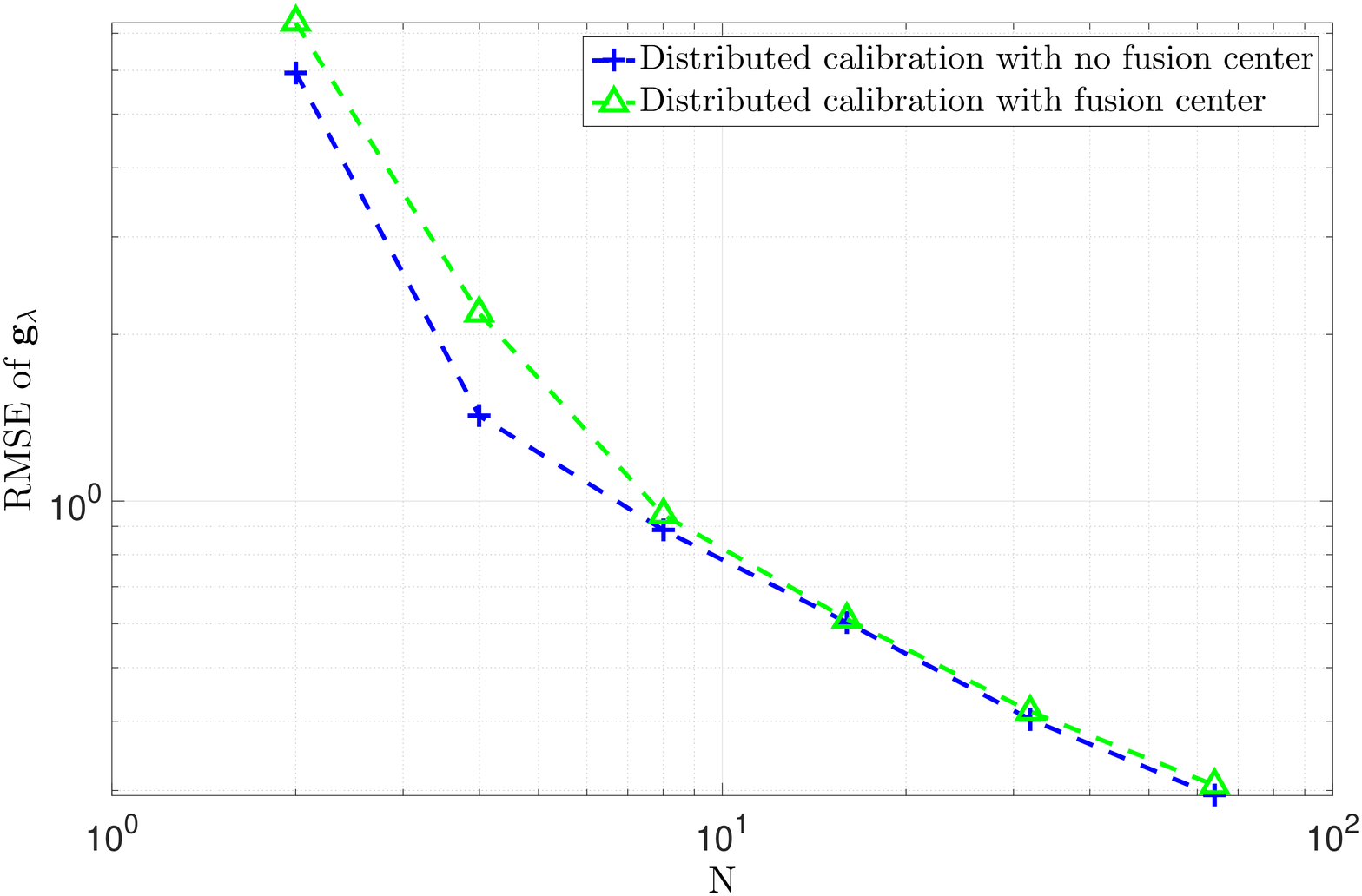}}
\caption*{Figure 3: (a) Statistical comparison between different network topologies for same computational cost, (b) Statistical comparison between different network topologies for different computational cost. }
\end{figure}

 \begin{figure}[t!]
 \centering
\subfigure[]{\label{fig:res2-a}\includegraphics[width=8cm, height=7cm]{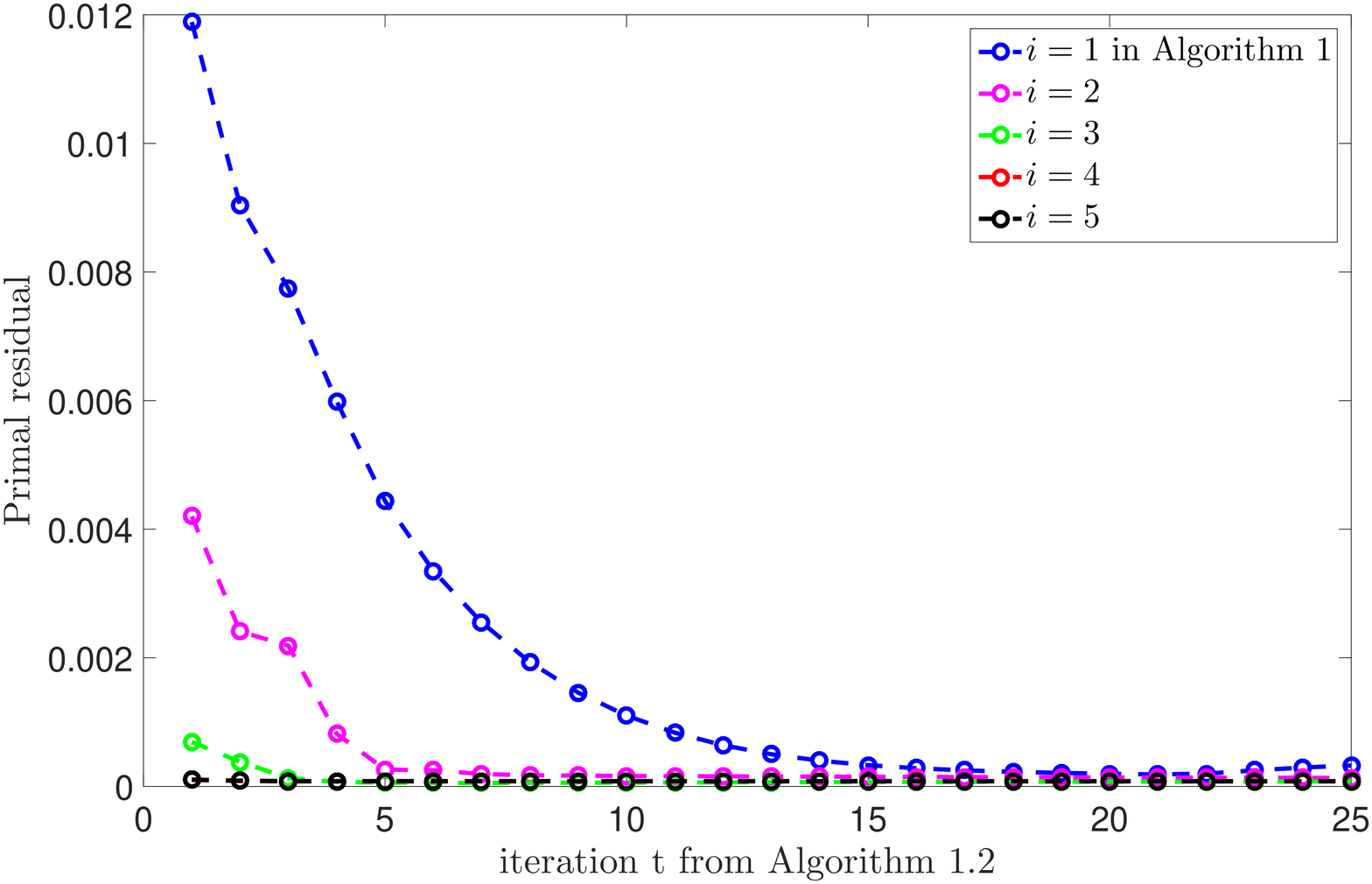}} \hspace{0.5pt} %
\subfigure[]{\label{fig:res2-b}\includegraphics[width=8cm, height=7cm]{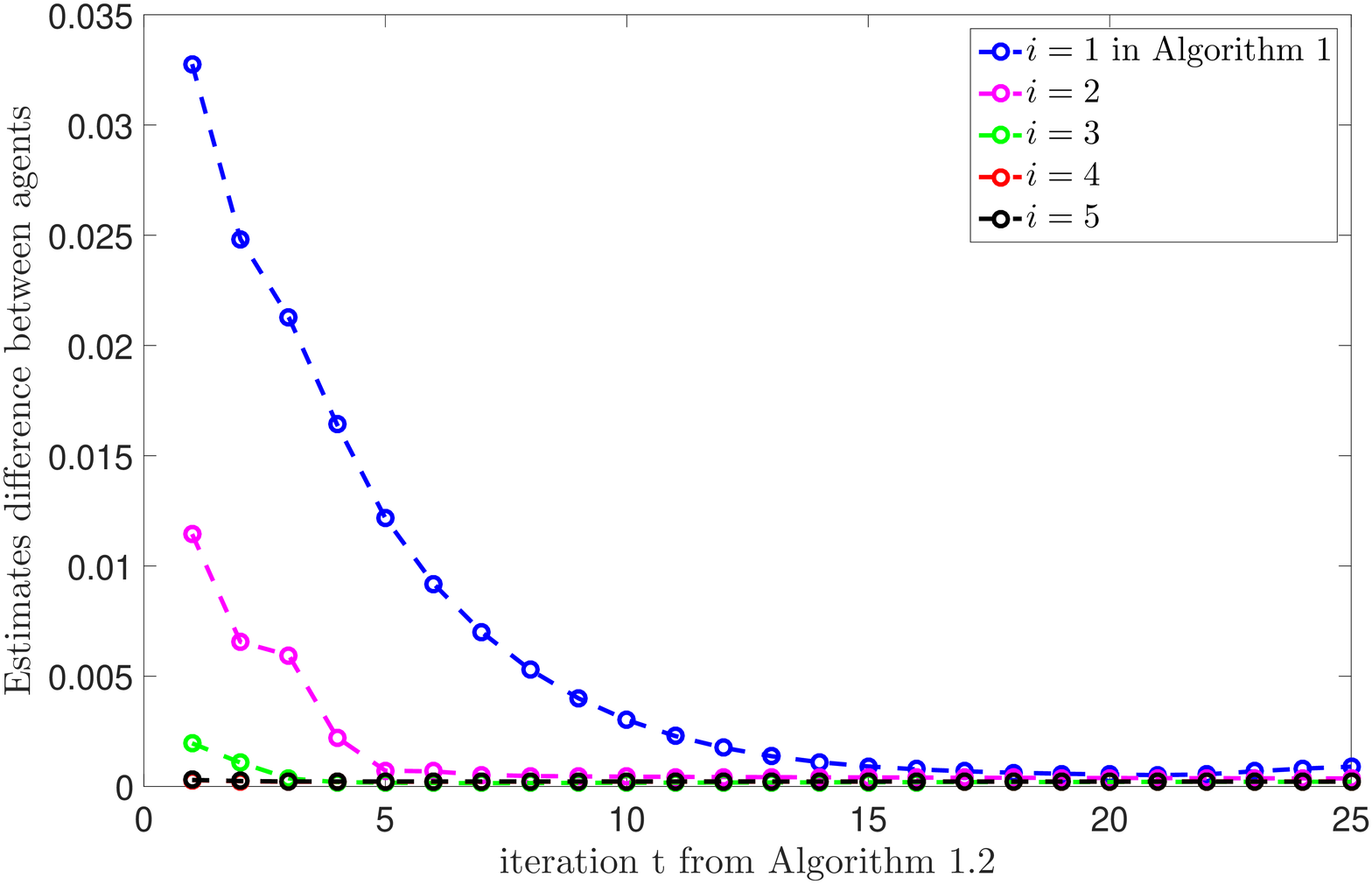}}
\caption*{Figure 4: (a) Primal residual $\epsilon_{p}$ and (b) estimates difference $\epsilon_{\mathrm{DIFF}}$ of the local consensus variables among agents as function of the iteration $t$ in Algorithm 2, for different values of the iteration $i$ in Algorithm 1. }
\end{figure}


\newpage

\bibliographystyle{elsarticle-num}
\bibliography{biblio}

\end{document}

%% file: new_commands.tex
\newcommand{\bfA}{\mathbf{A}}
\newcommand{\sfA}{\mathsf{A}}
\newcommand{\tbfA}{\tilde{\mathbf{A}}}

\newcommand{\bfb}{\mathbf{b}}

\newcommand{\bfB}{\mathbf{B}}
\newcommand{\calB}{\mathcal{B}}

\newcommand{\bbC}{\mathbb{C}}

\newcommand{\hbfd}{\hat{\mathbf{d}}}

\newcommand{\bfD}{\mathbf{D}}

\newcommand{\hbfD}{\hat{\mathbf{D}}}

\newcommand{\tbfE}{\tilde{\mathbf{E}}}

\newcommand{\calE}{\mathcal{E}}

\newcommand{\sfF}{\mathsf{F}}
\newcommand{\bfg}{\mathbf{g}}

\newcommand{\bfG}{\mathbf{G}}

\newcommand{\bfH}{\mathbf{H}}
\newcommand{\calH}{\mathcal{H}}
\newcommand{\sfH}{\mathsf{H}}

\newcommand{\bfI}{\mathbf{I}}

\newcommand{\tL}{\tilde{L}}

\newcommand{\bfm}{\mathbf{m}}
\newcommand{\tbfm}{\tilde{\mathbf{m}}}
\newcommand{\hbfm}{\hat{\mathbf{m}}}
\newcommand{\cm}{\check{m}}
\newcommand{\cbfm}{\check{\mathbf{m}}}

\newcommand{\bfM}{\mathbf{M}}
\newcommand{\tbfM}{\tilde{\mathbf{M}}}

\newcommand{\rmn}{\mathrm{n}}

\newcommand{\calN}{\mathcal{N}}
\newcommand{\bfp}{\mathbf{p}}

\newcommand{\hbfr}{\hat{\mathbf{r}}}
\newcommand{\chbfr}{\check{\hat{\mathbf{r}}}}
\newcommand{\bfR}{\mathbf{R}}
\newcommand{\bbR}{\mathbb{R}}
\newcommand{\hbfR}{\hat{\mathbf{R}}}

\newcommand{\calS}{\mathcal{S}}

\newcommand{\sfT}{\mathsf{T}}
\newcommand{\bfu}{\mathbf{u}}

\newcommand{\bfV}{\mathbf{V}}

\newcommand{\cbfV}{\check{\mathbf{V}}}
\newcommand{\bfx}{\mathbf{x}}

\newcommand{\bfy}{\mathbf{y}}

\newcommand{\bfz}{\mathbf{z}}
\newcommand{\bfZ}{\mathbf{Z}}
\newcommand{\tbfZ}{\tilde{\mathbf{Z}}}
\newcommand{\calZ}{\mathcal{Z}}
\newcommand{\boalpha}{\boldsymbol{\alpha}}

\newcommand{\hboalpha}{\hat{\boldsymbol{\alpha}}}

\newcommand{\bobeta}{\boldsymbol{\beta}}
\newcommand{\bogamma}{\boldsymbol{\gamma}}

\newcommand{\boGamma}{\boldsymbol{\Gamma}}

\newcommand{\boomega}{\boldsymbol{\omega}}
\newcommand{\boOmega}{\boldsymbol{\Omega}}

\newcommand{\bosigma}{\boldsymbol{\sigma}}
\newcommand{\boSigma}{\boldsymbol{\Sigma}}

\newcommand{\hbosigma}{\hat{\boldsymbol{\sigma}}}

\newcommand{\tboSigma}{\tilde{\boldsymbol{\Sigma}}}

\newcommand{\bfRzero}{\mathbf{R}^{\TRUE}}

\newcommand{\boSigman}{\boldsymbol{\Sigma}^{\mathrm{n}}}

\DeclareMathOperator{\diag}{diag}

\DeclareMathOperator{\vect}{vec}
\DeclareMathOperator{\vectdiag}{vecdiag}
\DeclareMathOperator{\blkdiag}{blkdiag}

\newcommand{\bfun}{\mathbf{1}}

\newcommand{\OUT}{\text{\tiny U}}
\newcommand{\TRUE}{\text{\tiny K}}

\newcommand{\RMSE}{\mathrm{RMSE}}

%% file: tikz_dessin.tex
\begin{tikzpicture}[scale=1,transform shape]
\def\rayon{4}
\pgfmathsetmacro\xdeux{-cos(60)*\rayon}
\pgfmathsetmacro\ydeux{sin(60)*\rayon}
\GraphInit[vstyle=Normal]
\renewcommand*{\VertexLineColor}{black}
\Vertex[x=-\rayon, y=0, L=$\sfA_1$]{A1}
\Vertex[x=\xdeux, y=\ydeux, L=$\sfA_2$]{A2}
\Vertex[x=-\xdeux, y=\ydeux, L=$\sfA_3$]{A3}
\Vertex[x=\rayon, y=0, L=$\sfA_4$]{A4}
\Vertex[x=-\xdeux, y=-\ydeux, L=$\sfA_5$]{A5}
\Vertex[x=\xdeux, y=-\ydeux, L=$\sfA_6$]{A6}

\tikzstyle{EdgeStyle}=[post]
   \Edge[label={$\bogamma^{1,2}$}, style={bend left=20}](A1)(A2)
   \Edge[label={$\bogamma^{2,1}$},style={bend left=20}](A2)(A1)
   \Edge[label={$\bogamma^{2,3}$},style={bend left=20}](A2)(A3)
   \Edge[label={$\bogamma^{3,2}$},style={bend left=20}](A3)(A2)
   \Edge[label={$\bogamma^{3,4}$},style={bend left=20}](A3)(A4)
   \Edge[label={$\bogamma^{4,3}$},style={bend left=20}](A4)(A3)
   \Edge[label={$\bogamma^{4,5}$},style={bend left=20}](A4)(A5)
   \Edge[label={$\bogamma^{5,4}$},style={bend left=20}](A5)(A4)
   \Edge[label={$\bogamma^{5,6}$},style={bend left=20}](A5)(A6)
   \Edge[label={$\bogamma^{6,5}$},style={bend left=20}](A6)(A5)
   \Edge[label={$\bogamma^{6,1}$},style={bend left=20}](A6)(A1)
   \Edge[label={$\bogamma^{1,6}$},style={bend left=20}](A1)(A6)
   
   \node[left] (A1) at (-\rayon-0.4, 0) {$\lbrace\hbfR_{\lambda}\rbrace_{\lambda \in \Lambda_1}$};
   \node[above] (A2) at (\xdeux, \ydeux+0.4) {$\lbrace\hbfR_{\lambda}\rbrace_{\lambda \in \Lambda_2}$};
   \node[above] (A3) at (-\xdeux, \ydeux+0.4) {$\lbrace\hbfR_{\lambda}\rbrace_{\lambda \in \Lambda_3}$};
   \node[right] (A4) at (\rayon+0.4, 0) {$\lbrace\hbfR_{\lambda}\rbrace_{\lambda \in \Lambda_4}$};
   \node[below] (A5) at (-\xdeux, -\ydeux-0.4) {$\lbrace\hbfR_{\lambda}\rbrace_{\lambda \in \Lambda_5}$};
   \node[below] (A6) at (\xdeux, -\ydeux-0.4) {$\lbrace\hbfR_{\lambda}\rbrace_{\lambda \in \Lambda_6}$};;

\end{tikzpicture}